# Two-dimensional metallic ferroelectricity in PbTe monolayer by electrostatic doping

Tao Xu,[1*] Jingtong Zhang[2], Yuquan Zhu[1], Jie Wang[2**], Takahiro Shimada[3], Takayuki Kitamura[3],

Tong-Yi Zhang[1***]

[1]*Materials Genome Institute, Shanghai University, Shanghai 200444, China,*

[2]*Department of Engineering Mechanics, School of Aeronautics and Astronautics, Zhejiang University, Hangzhou 310027, China,*

[3]*Department of Mechanical Engineering and Science, Kyoto University, Nishikyo-ku, Kyoto 615-8540, Japan*

**Abstract**

Polar metals characterized by the simultaneous coexistence of ferroelectric distortions and metallicity have attracted tremendous attention. Developing such materials at low dimensions remains challenging since both conducting electrons and reduced dimensions are supposed to quench ferroelectricity. Here, based on first-principles calculations, we report the discovery of ferroelectric behavior in two-dimensional (2D) metallic materials with electrostatic doping, even though ferroelectricity is unconventional at the atomic scale. We reveal that PbTe monolayer is intrinsic ferroelectrics with pronounced out-of-plane electric polarization originated from its non-centrosymmetric buckled structure. The ferroelectric distortions can be preserved with carriers doping in the ferroelectric monolayer, which thus enables the doped PbTe monolayer to act as a 2D polar metal. With an effective Hamiltonian extracted from the parametrized energy space, we found that the elastic-polar mode interaction is of great importance for the existence of robust polar instability in the doped system. The application of this doping strategy is not specific to the present crystal, but is rather general to other 2D ferroelectrics to bring about the fascinating metallic ferroelectric properties. Our findings thus change conventional acknowledge in 2D materials and will facilitate the development of multifunctional material in low dimensions.

**Introduction**

Polar metal with the coexistence of metallicity and polar structure is of significant interest in both fundamental science and application fields.[1-3] From the conventionally point of view, ferroelectricity can only exist in insulating or semiconducting materials because the static internal fields arising from dipolar orders are fully screened out by conduction electrons in metals. This common belief was challenged by the first experimental observation of a centrosymmetric-to-noncentrosymmetric structural transition in metallic $LiOsO_3$[4], which suggests the possibility of coexisting ferroelectricity and metallicity. Since then, enormous efforts have been devoted to identifying and designing new systems with these seemingly contra-indicated characteristics. For instance, polar metals were achieved by interface engineering in oxide heterostructures [5, 6]. Some other ferroelectric metals were also proposed by doping well-established ferroelectric materials, such as $BaTiO_3$[7] and $PbTiO_3$[8, 9].

With the fast-growing development of next-generation nanoelectronic devices, two-dimensional (2D) layered materials including graphene, silicene, phosphorene, and transition metal dichalcogenides have stimulated tremendous attention and research efforts.[10-16] They are endowed with a plethora of novel properties and physical phenomenon, and show promising potential in applications of nanoscale electronic and optoelectronic devices, and energy storage and conversion technologies.[17-20] Very recently, a few 2D materials have been experimentally demonstrated or theoretically proposed to have non-trivial intrinsic ferroelectricity[21-29], although the depolarization field is significantly enhanced at nanoscale and is believed to suppress electric polarization in common ferroelectrics. Despite the tremendous progresses, however, metallic ferroelectricity in atomistic 2D materials is rarely explored, since both conducting electrons and reduced dimensions are supposed to quench ferroelectricity. The discovery of new 2D ferroelectric metals and the underlying mechanism will not only unveil novel physics associated with them, but also promote the development of nanodevices.

Here, we predict that PbTe monolayer with electrostatic doping is a polar metal based on the first-principles calculations. Although PbTe monolayer has not been synthesized yet, a few atomic layer PbTe has been successfully grown in experiments [21]. Monolayer PbTe and its analogues have attracted a lot of theoretical attention for their fascinating physical properties [30-34]. We will demonstrate that the monolayer exhibits spontaneous polarization owing to its non-centrosymmetric buckled geometry. Moreover, the ferroelectric distortion shows a great compatibility with excess electrons, which can be introduced either intentionally or unintentionally in practice [35-38]. This leads to metallic ferroelectricity in the carriers doped PbTe monolayer.

The theoretical calculations were performed within the density functional theory (DFT) using the VASP package [39,40]. The Perdue–Burke–Ernzerhof (PBE) version of the generalized gradient approximation (GGA) was adopted as the exchange-correlation functional. The projected augmented wave (PAW) method was used for the electron-ion interactions with a plane wave cutoff energy of 400 eV. The Brillouin zone was sampled using a $15 \times 15 \times 1$ Monkhorst–Pack $k$-point mesh for the unit cell. A vacuum region of 30 Å normal to the monolayer was set to avoid any undesirable interactions between the neighboring monolayers due the periodic boundary conditions. The structures were fully relaxed until the energy change was less than $10^{-6}$ eV and the forces on each atom were less than 0.001 eV Å−1. The finite displacement method as implemented in the Phonopy package [41] was used to calculate the phonon spectrum of the PbTe monolayer. In addition, *Ab initio* molecular dynamics (AIMD) simulation with NVT ensemble and Nosé-Hoover thermostat were employed to assess the thermal stability of PbTe monolayer. Carriers doping was modeled by adding or removing electrons in the system compensated with uniform background charges. An effective Hamiltonian model was built to elucidate the energy contributions from different interactions that stabilize the ferroelectric phase. The details of the effective Hamiltonian can be found in the Supplementary Information.

The crystal structure of bulk PbTe is characterized by rocksalt-type structure with $Fm\bar{3}m$ space group. In its monolayer form, existing reports indicate that PbTe is in square shape with (space group *P4mm*) [33] or without (space group *Pmmm*) buckling [32], while recent high-throughput calculations [42] suggest that the buckled trigonal geometry (space group *P3m1*) is the stable structure. Here, our phonon dispersion calculations reveal that the buckled square and flat square structures are dynamically unstable because of the existence of soft mode in Brillouin zones (see Figure S1). Energetics of the buckled square and flat square structures are 0.25 eV/f.u. and 0.41 eV/f.u. higher than that of *P3m1* trigonal structure, respectively, which also indicates the *P3m1* trigonal structure to be the ground state structure. The schematic view of the trigonal structure after full relaxation is presented in Figure 1(a). In this geometry, PbTe monolayer consists of two atoms in a primitive unit cell and each Pb (Te) atom has a three-fold coordination with Te (Pb) atoms. The equilibrium in-plane lattice constants are $a_0 = b_0 = 4.33$ Å and the buckling value *u*, i.e. the displacement between the Pb and Te atoms along the (001) direction, is 1.69 Å, which are consistent with previous calculations [42].

The phonon spectrum of this structure is presented in Figure 1(b). It is evident that the studied monolayer is free from any imaginary-frequency modes within the entire Brillouin zone, which is a clear signature of its dynamic stability. We also calculate the cohesive energy of the PbTe monolayer $E_c = E_{PbTe} - E_{Pb} - E_{Te}$, with $E_{PbTe}$, $E_{Pb}$ and $E_{Te}$ being the total energies of unit cell of PbTe monolayer, isolated Pb atom and Te atom, respectively. The obtained cohesive energy is $-5.60$ eV, indicating the energetic favor of the formation of the monolayer configuration. In addition, the mechanical stability of the monolayer is assessed by calculating the elastic constants of the crystal. The derived independent elastic constants are: $C_{11}=5.8$ GPa, $C_{22}=5.8$ GPa, $C_{12}=0.2$ GPa, $C_{44}=2.1$ GPa. Obviously, the monolayer structure satisfies the mechanical stability of Born criteria of 2D materials [45], i.e. $C_{11}C_{22} - C_{12}^2 > 0$ and $C_{11}, C_{22}, C_{44} > 0$. It is also necessary to check whether the monolayer can retain stable structure at finite temperature. To this end, we perform AIMD simulations at a temperature of 300 K. As illustrated in Figure S2, total energy of the studied crystal shows slight fluctuation around a certain constant value and snapshot of the crystalline structure after 5 ps simulation shows that the monolayer remains intact. This further indicates the thermal stability of PbTe monolayer. These results demonstrate the stability and plausibility of the PbTe monolayer.

Having assessed the stability of PbTe monolayer, we proceed to analyze the ferroelectricity of the 2D crystal. As the PbTe monolayer consists of two kinds of atoms with different electronegativity, the atomic buckling of the monolayer breaks the spatial inversion symmetry. The lattice distortion that quantifies the symmetry breaking can be defined by the buckling parameter $u$, as denoted in Figure 2(a). This non-centrosymmetric structure (denoted as $F$ phase) hints the existence of out-of-plane electric polarization since ferroelectricity is derived from the separation of negative and positive charges. We also note that the buckled $F$ phase can transform into an energy degenerate configuration $F'$ with opposite buckling value via a spatial inversion operation. As a result, $F$ and $F'$ belong to the same ferroelectric phase. In comparison, when the buckling value of $\boldsymbol{u} = 0$, we have a centrosymmetric structure ($P$ phase) as illustrated in the middle of Figure 2(a), which forbids spontaneous polarization. This flat state with optimized in-plane lattice parameters of $a' = b' = 5.31$ Å is considered as the intermediate configuration between $F$ and $F'$ phase transformation.

Our Berry phase analysis indeed confirms the existence of ferroelectricity in the buckled monolayer. The calculated spontaneous polarization is equivalent to $29.2 \times 10^{-10}$ C m$^{-1}$ along [001] direction, while the in-plane components cancel out. This value is one order of magnitude larger than that in ultrathin ferroelectric SnTe ($\approx 10^{-10}$ C m$^{-1}$[21]) as has been detected in experiment. Polarization switching in this ferroelectric monolayer can be achieved by folding the direction of Pb-Te bonds. We further employed the solid-state nudge elastic band (SSNEB) method [22,23] to estimate the minimum energy pathway for polarization conversion. The results are presented in Figure 2(b), in which the initial and final states are set as $F$ and $F'$ phases, respectively. One can see that the energies of buckled structures are gradually increased with increasing buckling height and the energy barrier for the transition is estimated to be about 0.58 eV. This value is the same order as those in conventional FE oxides (e.g. 0.1 to 0.2 eV/f.u. for PbTiO$_3$) and other 2D ferroelectric materials. [46, 47] Furthermore, we apply a vertical electric field to the monolayer to directly inspect the feasibility of polarization reversal. With increasing applied electric field, the Pb atoms with positive charge gradually move down whereas the Te atoms with negative charge move up, leading to the decrease of electric polarization. The polarization direction is reverted suddenly when the electric field is larger than 30 V/nm, which is very close to that predicted in other 2D ferroelectrics [48]. All these results demonstrate that PbTe monolayer is intrinsic ferroelectrics with switchable polarization. Similar ferroelectric monolayers with out-of-plane spontaneous polarization have also been proposed very recently. [46−49]

We now turn to examine the effect of electrostatic doping on its ferroelectric properties by adding or removing electrons. The range of doping density investigated in the present study is up to 0.2 e/f.u., corresponding to the carrier density of $1.24 \times 10^{14}$ cm$_{-2}$. Such kind of electron modulation for 2D materials is experimental accessible by utilizing an electrolytic gate [43,44]. The variation of ferroelectric distortion in a response to carrier doping density is depicted in Figure 3(a), in which positive and negative carrier densities indicate the electron and hole doping, respectively. In conventional ferroelectrics, the excess electrons are believed to be detrimental to the electric polarization due to screening effect, thus leading to the disappearance of polar distortion for electron density beyond a certain critical value. Surprisingly, the polar distortion here in PbTe monolayer is seen to only be marginally affected by electron doping. Although the distortion slightly decreases with electron doping, as large as $d = 1.62$ Å still remains when the injected electron concentration is raised to a large density of 0.20 $e$/ f.u.. On the other hand, the polar distortion is slightly increased with increasing hole density within moderate doping level and then hardly change at high doping density. Our results thus reveal that there is no critical carrier concentration for the disappearance of ferroelectricity in the PbTe monolayer, and the carriers do not suppress but coexist with the ferroelectric properties within practical doping density.

The electronic properties of the doped crystal are also analyzed to clarify the characteristics of the doped carriers. Figure 3(b) presents the spin-resolved total density of states (DOS) of the electron doped PbTe monolayer at a concentration of $n = 0.10$ $e$/ f.u., in which the spin-up and spin-down states are completely degenerate. The pristine PbTe monolayer without carrier doping is a semiconductor with a band gap of about 1.5 eV (see Figure S3). Nevertheless, the doping electrons introduce new occupied states and lead to the valence and conduction bands overlap, which highlights a marked metallic character in the system. The projected DOS demonstrates that the weight at the Fermi level, i.e. the excess electrons, are primarily contributed by the $s$ and $p_z$ orbital of Pb atoms. This can also be visualized from their squared wave functions. These delocalized free electrons contribute to electrical conduction in the system. Similar metallic properties have also been observed in the PbTe monolayer with hole doping. Thus, in sharp contrast to conventional ferroelectrics such as $BaTiO_3$, where polar distortions and metallic conductivity are distinctly incompatible with each other, the ferroelectric instability unexpectedly persists with free carriers in atomic PbTe layer.

This unusual interaction between ferroelectric distortion and excess electrons can be understood from the energy contributions of different interactions in the stabilization of ferroelectric phase. We constructed an effective Hamiltonian that represents the energy change during paraelectric to ferroelectric phase transition. The parameterized Hamiltonian $E_{total}$ consists of the energy contributions from local mode self-energy $E_{self}$, short range interactions $E_{short}$, dipole-dipole interactions $E_{dpl}$, elastic-local mode interactions $E_{int}$ and elastic energy $E_{elas}$, i.e.

$$E^{tot}(\{u\},\eta) = E^{self}(\{u\}) + E^{short}(\{u\}) + E^{dpl}(\{u\}) + E^{int}(\{u\},\eta) + E^{elas}(\eta), \qquad (1)$$

where $u$ and $\eta$ are the local soft mode and local strain, respectively. See Supplementary Material for details on the expansion of each energy term. All the expansion parameters of the effective Hamiltonian for PbTe monolayer as well as perovskite oxide $BaTiO_3$ determined from DFT calculations are listed in Table S1 and Table S2. The estimated energy curves of different interactions in BaTiO$_3$ with and without doping are plotted in Figure S4. Ferroelectricity in BaTiO$_3$ originates from a delicate balance of all the energy contributions (see Figure S4(a)). Upon carrier doping, the magnitude of short-range interaction that contributes the most for lowering the total energy for ferroelectric transition is dramatically decreased due to the screening effect of the doped carriers (see Figure S4(b)). Therefore, polar distortions decrease with increasing doping density and eventually vanish at high density in BaTiO$_3$ (e.g. beyond 0.085 e/f.u. [50]). By contrast, we can find that the coupling between the elastic deformations and the local modes contributes greatly for lowering the total energy in the ferroelectric phase of the PbTe monolayer (see Figure 4(a)). This strain-local mode interaction drives large compressive strain and buckling atomic structure, and thus stabilizes its ferroelectric buckled phase. Remarkably, this energy contribution is insensitive to the doping effect and is not screened by doping as illustrated in Figure 4(b) for the electron doping density of $n = 0.1$ e/f.u.. As a result, the ferroelectric distortions persist in the doped PbTe monolayer.

The above results thus demonstrate the coexistence of polar instability and excess electrons in the PbTe monolayer due to the strong elastic-local mode interactions, which also provide a possibility in the search of new polar metals. Similar results can be anticipated for other 2D ferroelectric monolayers. Recent theoretical study predicted several 2D ferroelectric AB binary monolayers with out-of-plane polarization, including SiGe, GeSn and so forth.[49] We analyze the effect of electrostatic doping on polar distortions in these ferroelectric monolayers and find that polar distortions also resist metallization. Although different mechanisms (e.g. the lone-pair mechanism) [51] or strategies (e.g. geometric design [6]) have also been proposed to realize bulk polar metals, the present strategy extends the concept of polar metal to ultimately thin monolayers, which may have fascinating perspectives for nanoelectronics applications.

In conclusion, we predict that carriers doped PbTe monolayer is 2D metallic ferroelectricity by using first-principles calculations. The dynamical, thermal and mechanical stabilities of the PbTe monolayer are confirmed theoretically. The monolayer exhibits pronounced out-of-plane spontaneous polarization due to its non-centrosymmetric buckled structure. Remarkably, the ferroelectric distortions exhibit great compatibility with mobile carriers in such kind of ferroelectric monolayer attributed to the strong elastic-mode interaction. Our findings broaden the family of polar metal and lead to new types of multifunctional materials in low dimensions.

**Acknowledgements**

The authors acknowledge the financial support from the National Natural Science Foundation of China (Grant No. 11802169).

**Author Information**

**Corresponding author.**

\*   E-mail: xutao6313@shu.edu.cn
\*\*  E-mail: jw@zju.edu.cn
\*\*\* E-mail: zhangty@shu.edu.cn

**Author Contributions.** T.X. conceived the project, designed and directed computational experiments, and wrote the entire manuscript. J.T.Z. built the effective Hamiltonian model and discussed the results. Y.Q.Z supported the calculations. T. S. and T. K. discussed the results. J.W. and T.Y.Z. supervised the work and provided critical feedback on the manuscript. All authors read and commented on the manuscript. T.X. and J.T.Z. contributed equally to this work.

**Notes**

The authors declare no competing financial interest.

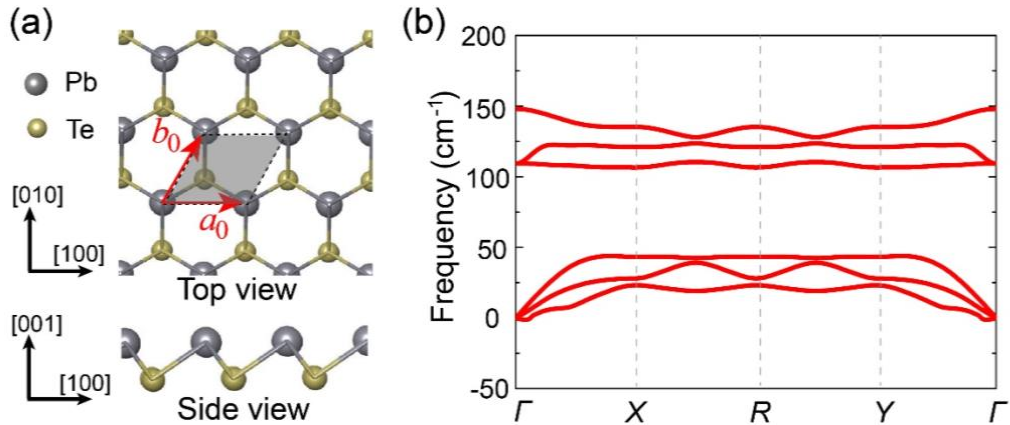

**Figure 1.** (a) Top view and side view of PbTe monolayer. The shaded rhombus denotes the primitive cell, in which the two basis vectors are denoted as $a_0$ and $b_0$. (b) Phonon dispersion curves of PbTe monolayer.

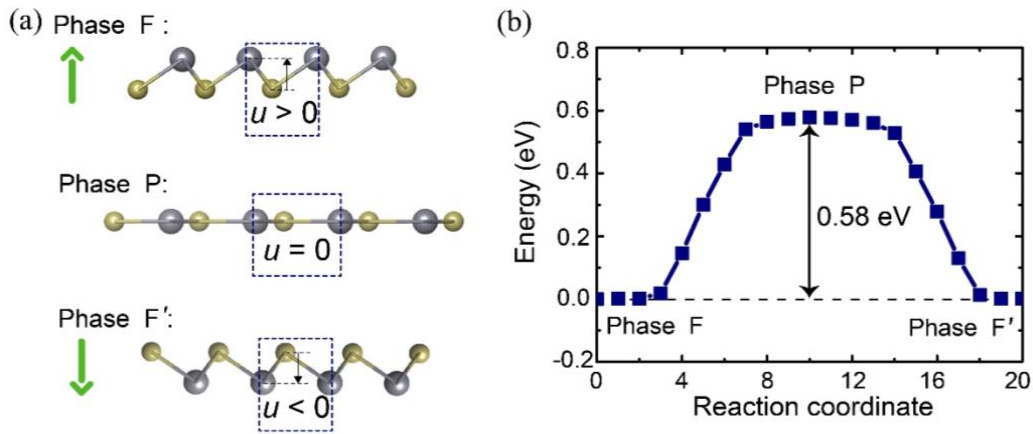

**Figure 2.** (a) Side view of the atomic structures of two energy-degenerated distorted phases (F and F´) and the centre-symmetric P phase; The green arrows indicate the polarization directions. (b) Total energy change for the transformation between two ferroelectric phases (F and F´) in the PbTe monolayer obtained from SSNEB method. The reaction coordinate number 0 and 20 denote ferroelectric phases F and F´, respectively.

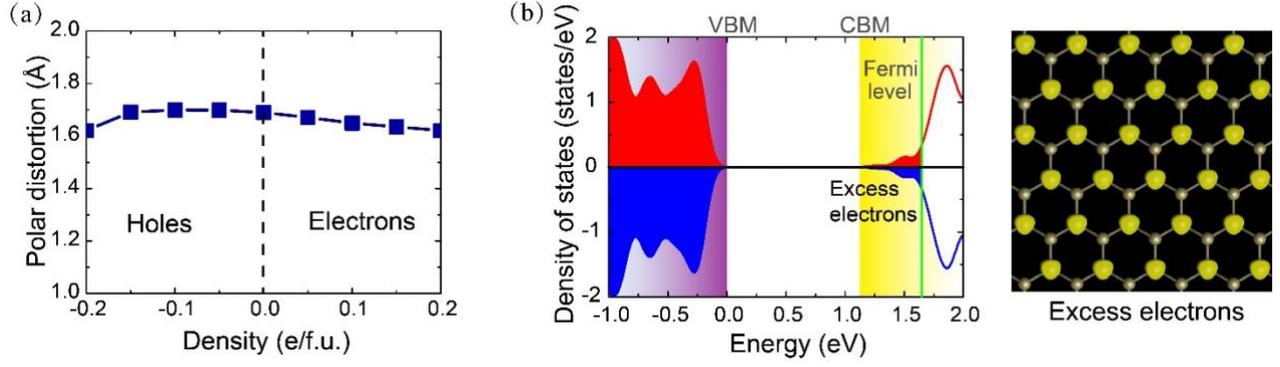

**Figure 3.** (a) Evolution of polar distortions with electrostatic doping; positive and negative densities indicate the electron and hole doping, respectively. (b) Spin-resolved total DOS of the electron doped PbTe monolayer at a concentration of $n = 0.10$ $e$/ f.u. The green solid line denotes the Fermi level. The corresponding squared wave functions of the excess electrons are shown in the right panel of (b).

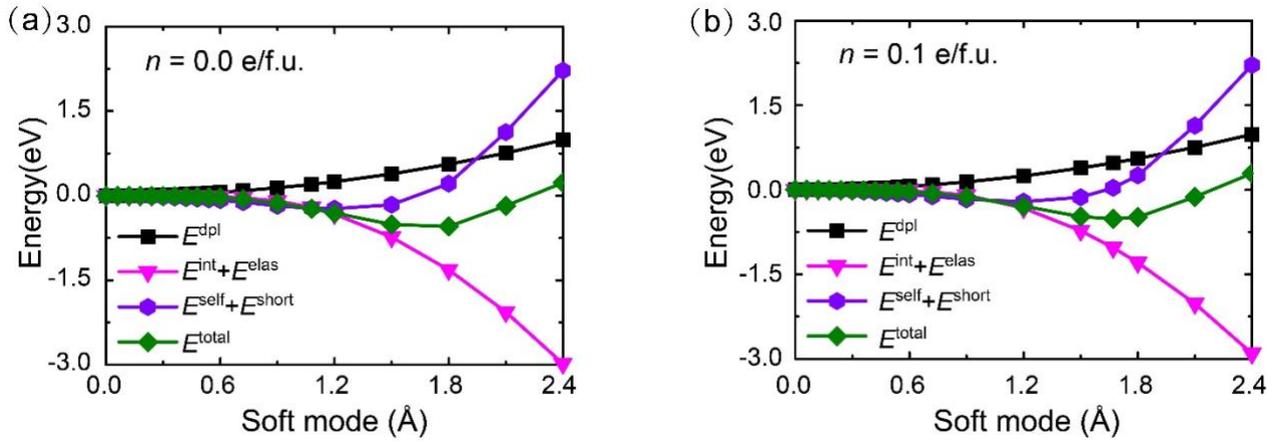

**Figure 4.** Energy contributions in the Hamiltonian during the ferroelectric phase transition in the PbTe monolayer at densities of (a) $n =0$ (b) $n = 0.10$ $e$/ f.u..

17